\documentclass[%
 reprint,
superscriptaddress,
frontmatterverbose, 
nofootinbib,
 amsmath,amssymb,
 aps,
prb,
]{revtex4-2}
\usepackage[utf8]{inputenc}
\usepackage{amsmath}
\usepackage{amsfonts}
\usepackage{amssymb}
\usepackage{graphicx}
\usepackage{hyperref}
\usepackage{braket}
\usepackage[colorinlistoftodos]{todonotes}
\usepackage[left=2cm,right=2cm,top=2cm,bottom=2cm]{geometry}
\usepackage{cleveref}

\usepackage[caption=false]{subfig}
\usepackage[T1]{fontenc}

\begin{document}

\title{Noise cross-correlations from single-shot measurements}

\author{J. S. Rojas-Arias}
\email{juan.rojasarias@riken.jp}
\affiliation{RIKEN, Center for Quantum Computing (RQC), Wako-shi, Saitama 351-0198, Japan}
\author{P. Stano}
\affiliation{RIKEN, Center for Quantum Computing (RQC), Wako-shi, Saitama 351-0198, Japan}
\affiliation{Slovak Academy of Sciences, Institute of Physics, 845 11 Bratislava, Slovakia}
\author{Y.-H. Wu}
\affiliation{RIKEN, Center for Emergent Matter Science (CEMS), Wako-shi, Saitama 351-0198, Japan}
\affiliation{Department of Physics, National Taiwan University, Taipei 10617, Taiwan}
\author{L. C. Camenzind}
\affiliation{RIKEN, Center for Emergent Matter Science (CEMS), Wako-shi, Saitama 351-0198, Japan}
\author{S. Tarucha}
\affiliation{RIKEN, Center for Quantum Computing (RQC), Wako-shi, Saitama 351-0198, Japan}
\affiliation{RIKEN, Center for Emergent Matter Science (CEMS), Wako-shi, Saitama 351-0198, Japan}
\author{D. Loss}
\email{daniel.loss@unibas.ch}
\affiliation{RIKEN, Center for Quantum Computing (RQC), Wako-shi, Saitama 351-0198, Japan}
\affiliation{Department of Physics, University of Basel, Klingelbergstrasse 82, CH-4056 Basel, Switzerland}

\begin{abstract}

We introduce a novel method that we call Single-Shot Cross-Spectroscopy (SSCS), for extracting the auto- and cross-power spectral densities of dephasing noise of a qubit pair. The method uses straightforward input, namely single-shot readouts from single-qubit Ramsey-type experiments, and is resilient against errors in state preparation and measurement. We apply it to experimental data from a semiconductor spin-qubit device and obtain noise spectra over five orders of magnitude in frequency (5 mHz--500 Hz). Compared to other techniques, SSCS enables access to noise correlations in the previously inaccessible intermediate-frequency range (1--500 Hz) for spin qubits, and can be further extended with faster readout. More broadly, the frequency range accessible with SSCS is limited only by the experiment repetition rate, and scales accordingly on other platforms.
\end{abstract}

\maketitle

\section{Introduction}

Understanding and mitigating noise is a central challenge in quantum information processing. Apart from autocorrelations, which characterize the performance of isolated qubits, correlations in noise across different qubits also play a role in multi-qubit circuits. For example, correlated errors are problematic for quantum error correction, limiting the achievable improvements in logical error rates upon increasing the code distance~\cite{Google2023,Google2025}. Identifying and quantifying noise correlations is thus essential for advancing quantum architectures~\cite{Clemens2004,Boter2020}. Particularly in solid-state platforms, spatially correlated noise is ubiquitous, arising from common environmental~\cite{Yoneda2023,Rojas-Arias2023} or control-related sources~\cite{Undseth2023,Heinz2021,Zhao2022}.

One way to characterize noise is by reconstructing its spectral properties, an approach called noise spectroscopy. Naturally, different methods are suitable for detecting and quantifying noise at different frequencies. Since each method has its practical limitations, the noise spectra extracted from experiments often display gaps, ranges of frequencies that are difficult to access. 

Low-frequency noise ($f \lesssim 10$ Hz), 
%typically arising from slow fluctuations such as charge or hyperfine noise, 
can often be observed directly as qubit-energy shifts that can be tracked in time, repeatedly estimating the qubit energy from Ramsey measurements \cite{Yoneda2018,Ramsey1950}. The Fourier transform of these time traces gives the auto-power spectral density (auto-PSD)~\cite{Burnett2019,Vepsalainen2022,Yan2012,Delbecq2016,Nakajima2020,Struck2020,Malinowski2017}. The upper frequency limit of this approach is fundamentally set by the Nyquist frequency, which in turn depends on how fast a single estimate of the qubit energy can be made. A larger Nyquist frequency thus requires a shorter Ramsey cycle, which is practically limited by factors such as qubit initialization and readout times.

High-frequency noise ($f \gtrsim 10^5$ Hz) is accessed through dynamical decoupling sequences such as CPMG. They act as bandpass filters that amplify a selected narrow frequency range~\cite{Cywinski2008,Yan2013,Yoneda2018,Jock2022,Connors2022,Bylander2011}. On the lower frequency end, the CPMG is limited by the exponential decay of the signal due to decoherence upon prolonging the total sequence length. It typically happens at frequencies much higher than those accessible from time-tracked qubit energies, leaving a gap of a few or even many decades in frequency~\cite{Yoneda2018,Jock2022,Connors2022,Nakajima2020,Rojas-Arias2024,Fink2013}.

Concerning cross correlations, the situation is even more challenging. These can be quantified via the cross-power spectral density (cross-PSD), which captures frequency-resolved correlations between fluctuations in different qubits. At low frequencies, the cross-PSD can be extracted analogously to the auto-PSD, by tracking the qubit energies of multiple qubits simultaneously and computing the Fourier transform of their correlated fluctuations~\cite{Yoneda2023,Rojas-Arias2023,Rojas-Arias2024}. However, multi-qubit dynamical decoupling protocols, which would enable access to high-frequency cross correlations, are significantly more involved~\cite{Szankowski2016} and have not yet been demonstrated experimentally. In any case, the intermediate-frequency gap where neither of the methods applies remains. 

Here we introduce a spectroscopy method that delivers PSDs, both auto and cross, in a wide frequency range,  spanning from millihertz to nearly a kilohertz in our demonstration. For spin qubits, this extends the accessible spectrum by two orders of magnitude toward higher frequencies, thereby bridging the gap discussed above. Inspired by the proposal from Refs.~\cite{Fink2013,Sakuldee2020}, we employ single-shot measurements rather than estimating qubit energies from multiple Ramsey experiments, which extends the Nyquist frequency and allows for resolution of higher-frequency noise components. We refer to this technique as Single-Shot Cross-Spectroscopy (SSCS). We validate SSCS on both simulated and experimental data, demonstrating its practical suitability for integration into standard noise characterization workflows in multi-qubit systems. To facilitate its adoption, we provide an open-source Python implementation of the full SSCS procedure~\cite{Notebook}.

\section{The method description}
\label{sec:method}

Below, we first describe the experimental sequences to produce data that is required as SSCS's input. We shed light on the method's essence by deriving the relation between the statistics of such data and the qubit-energy auto- and cross-PSDs. Finally, we validate the method on both simulated and experimental data.

\subsection{The required experimental sequences}
\label{sec:sequences}

\begin{figure}
\centering
\subfloat{\includegraphics[width=\columnwidth]{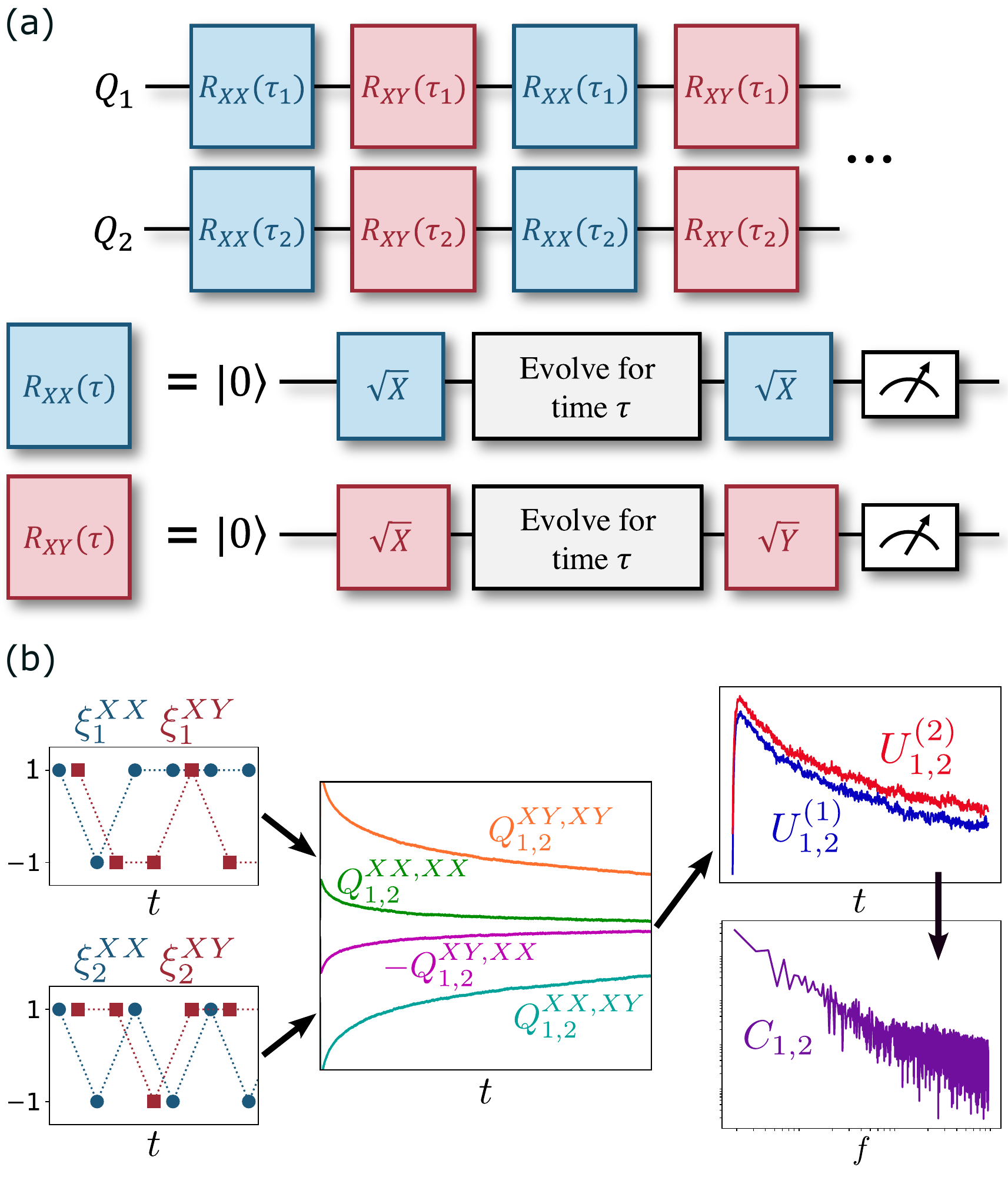}\label{fig:sequence}}
\subfloat{\label{fig:flow}}
\caption{(a) The proposed SSCS sequence acting on a qubit pair $Q_1-Q_2$ is shown in the upper panel, with the two subsequences $R_{XX}$ and $R_{XY}$ in the lower panel. (b) Procedure for extracting the cross-PSD of a qubit pair. The single-shot outcomes $\xi_\alpha^{if}$ from the SSCS sequence shown in (a) are used to compute the correlators $Q_{1,2}^{if,jh}$. These are then combined, using Eq.~\eqref{eq:UV}, to obtain the two $U^{(a)}_{1,2}$. A Fourier transform and average of these quantities yields the cross-PSD $C_{1,2}$.}
\label{fig:sequence_full}
\end{figure}

We consider two qubits, $Q_1$ and $Q_2$, whose qubit-frequency cross-PSD is of interest. SSCS requires performing single-qubit Ramsey sequences on each qubit---no two-qubit gates are needed. Therefore, the qubits may be spatially separated, but the single-qubit manipulations need to be synchronized with the structure shown in Fig.~\ref{fig:sequence}. Each qubit undergoes two alternating Ramsey subsequences, denoted $R_{XX}$ and $R_{XY}$. Here, $R_{if}$ represents a Ramsey experiment beginning with initialization to the $\ket{0}$ state, followed by a $\pi/2$ rotation about the $i$-axis, free evolution for a fixed time $\tau_\alpha$ (which may differ between qubits, $\tau_1 \neq \tau_2$), then a $\pi/2$ pulse about the $f$-axis, and a final projective single-shot readout in the $\ket{0}$ and $\ket{1}$ basis. The readout result is denoted by $\xi^{if}_\alpha$ and the two possible results should be encoded by $\xi \in \{ +1, -1\}$. The first $\pi/2$ pulse axis is always $i=X$, while the second pulse alternates between $f=X$ and $f=Y$. When $\tau_1 \neq \tau_2$, the second $\pi/2$ pulses may not be simultaneous across qubits; this does not affect the method. Moreover, the start times for the qubits only need to be aligned at the scale of the sequence repetition period $\Delta t$, typically micro- or milliseconds. There are a total $N$ of repetitions of the pairs $R_{XX}$--$R_{XY}$ of Ramsey cycles run on each qubit in total.

With the probing sequences defined, we next turn to the analysis of the measurement statistics. A detailed discussion of optimal choices for the sequence parameters is deferred to Sec.~\ref{sec:parameters}, where their impact can be more readily understood. At this stage, we focus on the set of single-shot measurement outcomes, denoted $\{ \xi^{if}_\alpha(t_n) \}_{\alpha,n}$, where $\alpha \in \{1,2\}$ labels the qubit, and $i,f \in \{X,Y\}$ label the axes of the $\pi/2$ pulses in each Ramsey subsequence. The integer index $n \in \{0,1,\dots,N-1\}$ labels time steps spaced by $2\Delta t$. The corresponding measurement times are given by $t^{if}_n = (2n + 1 - \delta_{if})\Delta t$, where $\delta_{if}$ is a Kronecker delta that offsets the timing to account for the interleaving of $R_{XX}$ and $R_{XY}$ sequences.

\subsection{Statistics of the single-shot results}
\label{sec:statistics}

The quantum-mechanical expectation value of a single-shot measurement $\xi^{if}_\alpha(t)$ of qubit $\alpha$ for the subsequence $R_{if}$ performed at laboratory time $t$ is given by
\begin{align}
P^{if}_\alpha(t) = A_\alpha+B_\alpha\sin[(\omega_\alpha+\delta\omega_\alpha(t))\tau_\alpha + \theta_{if}],
\label{eq:expectation}
\end{align}
where $\omega_\alpha$ is the average energy of the qubit, $A_\alpha$ and $B_\alpha$ account for state preparation and measurement (SPAM) errors, and we introduced the conditional phase $\theta_{if}=\frac{\pi}{2}\delta_{if}$. We assume the zero-mean\footnote{The mean does not need to be known exactly. A non-zero mean of $\langle \delta \omega_\alpha \rangle$ can be interpreted as an unknown offset in $\omega_\alpha$. Such an offset is not harmful as long as it is small, meaning $\langle \delta \omega_\alpha \rangle \tau_\alpha\ll 1$.} fluctuations of qubit energies $\delta\omega_\alpha$ are wide-sense stationary\footnote{Wide-sense stationary means that $\langle \delta\omega_\alpha(t)\rangle$ is independent of $t$, and the autocorrelation $\langle \delta\omega_\alpha(t)\delta\omega_\alpha(t+\tau)\rangle$ depends only on $\tau$.}, quasi-static\footnote{By quasi-static we mean that the noise is essentially constant during a single experiment run but varies randomly between runs.}, and Gaussian. The Gaussian-noise assumption is needed to evaluate ensemble averages, denoted as $\braket{\ldots}$ below. We adopt the quasi-static assumption to simplify the derivation and later give a generalized result which does not need it. 

Next, we perform the statistical average (the average over the noise realizations), using the identity $\braket{e^{i \phi}}=e^{-\braket{\phi^2}/2}$ for a Gaussian random variable $\phi$ with zero mean~\cite{Kubo1962}. The single-shot outcomes follow as
\begin{align}
\braket{P_\alpha^{if}}=A_\alpha+B_\alpha\sin(\phi_\alpha^{if})e^{-\tau_\alpha^2/T_{2,\alpha}^{*2}},
\label{eq:average}
\end{align}
where we have defined the coherence time of qubit $\alpha$ as $T_{2,\alpha}^*\equiv \sqrt{2/\braket{\delta\omega_\alpha^2}}$, and the phases $\phi_\alpha^{if}\equiv\omega_\alpha\tau_\alpha+\theta_{if}$.
Our main tool are the correlators of the single-shot readouts, 
\begin{align}
Q_{\alpha,\beta}^{if,jh}(t)\equiv\braket{P_\alpha^{if}(t^\prime)P_\beta^{jh}(t'+t)}-\braket{P_\alpha^{if}}\braket{P_\beta^{jh}}.
\label{eq:correlatorsOfP}
\end{align}
They can be evaluated analogously, through straightforward algebra, as
\begin{align}
&Q_{\alpha,\beta}^{if,jh}(t)=\frac{B_\alpha B_\beta}{2}\Big[\cos\left(\phi_\alpha^{if}-\phi_\beta^{jh}\right)e^{-\frac{\chi_{\alpha,\beta}^{(-)}}{2}} \nonumber\\
&\qquad \qquad \qquad \,\,
-\cos\left(\phi_\alpha^{if}+\phi_\beta^{jh}\right)e^{-\frac{\chi_{\alpha,\beta}^{(+)}}{2}} \label{eq:correlators}\\
&\qquad \qquad \qquad \,\, -2\sin\left(\phi_\alpha^{if}\right)\sin\left(\phi_\beta^{jh}\right) e^{-\left(\frac{\tau_\alpha^2}{{T}_{2\alpha}^{*2}}+\frac{\tau_\beta^2}{{T}_{2\beta}^{*2}}\right)} \Big],\nonumber
\end{align}
where the attenuation of the correlators is described by the following (``attenuation'') functions
\begin{align}
\begin{split}
\chi_{\alpha,\beta}^{\left(\pm\right)}(t) \equiv &\tau_\alpha^2\braket{\delta\omega_\alpha^2}+\tau_\beta^2\braket{\delta\omega_\beta^2}\\
&\pm2\tau_\alpha\tau_\beta\braket{\delta\omega_\alpha(t')\delta\omega_\beta(t'+t)}.
\end{split}
\label{eq:envelope}
\end{align}
Equations \eqref{eq:correlators} and \eqref{eq:envelope} are central to the understanding of the relation between the experimental sequences shown in Fig.~\ref{fig:sequence} and the PSDs that we aim to extract. 

We estimate the expectation values $\braket{P}$ and the correlators $Q$ from the measured single-shot readouts $\xi$ using the empirical estimators
\begin{align}
\braket{P_\alpha^{if}} =& \frac{1}{N}\sum_{n=0}^{N-1} \xi_\alpha^{if}(t_n), \\
Q_{\alpha,\beta}^{if,jh}(t_k) =& \frac{1}{N-k} \sum_{n=0}^{N-1-k} \xi_\alpha^{if}(t^{if}_n) \, \xi_\beta^{jh}(t^{jh}_{n+k}) \nonumber \\
& - \braket{P_\alpha^{if}} \, \braket{P_\beta^{jh}}.
\label{eq:estimators}
\end{align}
Here, $k \in \{0,1,\ldots,N-1\}$ is the time-lag index, and the corresponding delay is given by $t_k \equiv t^{jh}_{n+k} - t^{if}_{n} = (2k + \delta_{if} - \delta_{jh})\Delta t$.

\subsection{Relation of single-shot statistics to PSDs}
\label{sec:PSDs}

Our goal is to extract the time correlator of qubit energies $\braket{\delta\omega_\alpha(t')\delta\omega_\beta(t'+t)}$ from the attenuation functions in Eq.~\eqref{eq:envelope} such that we can obtain the PSD of the noisy qubit energies $C_{\alpha,\beta}(f)$ from its definition:
\begin{align}
C_{\alpha,\beta}(f)=\dfrac{1}{4\pi^2}\int_{-\infty}^\infty \mathrm{d} t\braket{\delta\omega_\alpha(t')\delta\omega_\beta(t'+t)}e^{2\pi i ft}.
\label{eq:cross-PSD def}
\end{align}
To do so, we note that the correlators $Q$ have a special structure, not immediately obvious,\footnote{The structure is in the prefactors of the three terms in Eq.~\eqref{eq:correlators}, related by trigonometric identities.} that allows one to isolate the time correlator $\braket{\delta\omega_\alpha(t')\delta\omega_\beta(t'+t)}$ exactly. There are two simple combinations that will do,\footnote{The quantity $U^{(a)}$ is a logarithm of a real number. However, the number might be negative. In this case, not to complicate the notation, we accept complex $U^{(a)}$, which will be a logarithm of a positive number plus $\pi i$. In Eq.~\eqref{eq:cross_UV}, the imaginary constant is absorbed in $c_a$ and the correlator $\braket{\delta\omega_\alpha(t')\delta\omega_\beta(t'+t)}$) remains real, as it should be.}
\begin{subequations}
\begin{align}
U_{\alpha,\beta}^{(1)}(t)&\equiv\log\left[\dfrac{Q_{\alpha,\beta}^{XX,XX}(t)+Q_{\alpha,\beta}^{XY,XY}(t)}{Q_{\alpha,\beta}^{XX,XX}(t)-Q_{\alpha,\beta}^{XY,XY}(t)}\right],\label{eq:U}\\
U_{\alpha,\beta}^{(2)}(t)&\equiv\log\left[\dfrac{Q_{\alpha,\beta}^{XY,XX}(t)-Q_{\alpha,\beta}^{XX,XY}(t)}{Q_{\alpha,\beta}^{XY,XX}(t)+Q_{\alpha,\beta}^{XX,XY}(t)}\right].\label{eq:V}
\end{align}%
\label{eq:UV}%
\end{subequations}
They both give
\begin{align}
U_{\alpha,\beta}^{(a)}(t)=\tau_\alpha\tau_\beta\braket{\delta\omega_\alpha(t')\delta\omega_\beta(t'+t)}+c_a,
\label{eq:cross_UV}
\end{align}
albeit the additive constants are different, $c_1 \neq c_2$.

Following Eq.~\eqref{eq:cross-PSD def}, the PSD is obtained with a Fourier transform (here $a=1$ or 2)
\begin{align}
C_{\alpha,\beta}(f)&=\frac{1}{4\pi^2\tau_\alpha\tau_\beta}\int_{-\infty}^\infty \mathrm{d}t\ U_{\alpha,\beta}^{(a)}(t)e^{2\pi i ft}+c_a \delta(f).
\label{eq:cross_single_shot}
\end{align}
with $\delta(f)$ a Dirac delta distribution. This is the final result that will be used to convert the single-shot results into noise PSDs. Next, we comment on some of its aspects.

\subsection{Discussion of the main result, Eq.~\eqref{eq:cross_single_shot}}
\label{sec:discusssionOfMainEq}

We begin by emphasizing a key advantage of SSCS: its robustness against SPAM errors. {This robustness results from two steps in the method.} First, the subtraction of the product of averages (the second term in Eq.~\eqref{eq:correlatorsOfP}) eliminates the bias introduced by SPAM errors characterized by the parameters $A$. As a result, the correlators $Q$ depend only on the visibility parameters $B$, which appear as prefactors in Eq.~\eqref{eq:correlators}. {Second, taking logarithms in the definition of $U^{(a)}$ transforms these prefactors into additive constants in Eq.~\eqref{eq:cross_UV}, which then appear as zero-frequency components after the Fourier transform in Eq.~\eqref{eq:cross_single_shot}.} {This insensitivity to SPAM errors is essential for any method intended for use with experimental data.}

Next, we note that Eq.~\eqref{eq:UV} provides two alternatives, yielding two independent estimates of PSD that can be combined to improve precision. Interestingly, the correct procedure is to perform the Fourier transform first and only then average the resulting spectra, rather than averaging $U^{(1)}$ and $U^{(2)}$ before applying Eq.~\eqref{eq:cross_single_shot}. The reason is that the two $U^{(a)}$ are evaluated at different time steps, due to the alternating application of $R_{XX}$ and $R_{XY}$ subsequences, spaced by $\Delta t$. Specifically, the measurement outcomes $\xi_\alpha^{XX}$ are recorded at times $2n\Delta t$, while the outcomes $\xi_\alpha^{XY}$ are recorded at $(2n + 1)\Delta t$, for $n = 0,\ldots,N-1$. This alternation propagates through the correlators $Q$ to the quantities $U^{(a)}$, yielding $U^{(1)}(2n\Delta t)$ and $U^{(2)}((2n+1)\Delta t)$. While the estimated cross-PSD is ultimately evaluated at the same set of frequencies, this temporal offset must be accounted for when computing the Fourier transform in Eq.~\eqref{eq:cross_single_shot}. As we show in Appendix.~\ref{app:aliasing}, the average of the estimated PSDs is much more robust to aliasing effects than the individual terms.

The SSCS method can also be applied to estimate auto-PSDs by setting $\alpha = \beta$ in Eq.~\eqref{eq:cross_single_shot}, albeit with a few caveats. First, the sequence parameters must be chosen such that $\cos(2\phi_\alpha^{if}) \neq 0$. This ensures that $Q_{\alpha,\alpha}^{XX,XX}(t) - Q_{\alpha,\alpha}^{XY,XY}(t) \neq 0$, {enabling the calculation of the auto-PSD $S_\alpha(f) \equiv C_{\alpha,\alpha}(f)$ from the Fourier transform of $U_{\alpha,\alpha}^{(1)}(t)$}. Second, since $Q_{\alpha,\alpha}^{XY,XX}(t) - Q_{\alpha,\alpha}^{XX,XY}(t) = 0$, Eq.~\eqref{eq:V} cannot be used to {reduce aliasing}. {Third, estimating $Q_{\alpha,\alpha}^{if,if}(t)$ at $t=0$ is problematic due to the discrete nature of the single-shot outcomes and the impossibility of measuring two outcomes with zero time delay. As shown in Eq.~\eqref{eq:estimators}, this estimate becomes $Q_{\alpha,\alpha}^{if,if}(0) = 1 - \braket{P_\alpha^{if}}^2$ due to $(\xi_\alpha^{if})^2 = 1$, whereas the correct expression from Eq.~\eqref{eq:correlatorsOfP} is $Q_{\alpha,\alpha}^{if,if}(0) = \braket{(P_\alpha^{if})^2} - \braket{P_\alpha^{if}}^2$.} Whatever value is adopted for it,\footnote{We extrapolate $Q_{\alpha,\alpha}^{if,if}(t)$ at $t=0$ from a fit to its values at $t>0$.} any deviation from the (unknown) true value of $Q_{\alpha,\alpha}^{if,if}(t=0)$ appears as a Dirac delta contribution $\delta(t)$ in $U^{(1)}_{\alpha,\alpha}(t)$, which becomes a white noise floor in $S_\alpha(f)$ after taking the Fourier transform. Finally, as mentioned in Refs.~\cite{Fink2013, Sakuldee2020,Rojas-Arias2024}, for evolution times longer than $T_2^*$ the attenuation functions fulfill $\chi_{\alpha,\alpha}^{(+)}\gg\chi_{\alpha,\alpha}^{(-)}$, such that in $Q_{\alpha,\alpha}^{if,if}$ only the terms with $\exp(-\chi_{\alpha,\alpha}^{(-)}(t)/2)$ survive. In this limit, it is possible to obtain the auto-PSD also as
\begin{align}
S_\alpha(f) = \frac{1}{4\pi^2 \tau_\alpha^2} \int_{-\infty}^{\infty} \mathrm{d}t\, W_\alpha(t) e^{2\pi i f t} + \text{const.} \times \delta(f),
\label{eq:auto-PSD}
\end{align}
where {$W_\alpha(t) \equiv \log\left[Q_{\alpha,\alpha}^{XX,XX}(t) + Q_{\alpha,\alpha}^{XY,XY}(t)\right]$}.

From our experience, we find Eq.~\eqref{eq:auto-PSD} to provide the better signal-to-noise ratio (SNR) when extracting auto-PSDs, compared to Eq.~\eqref{eq:cross_single_shot}. This is because for auto-PSDs the denominator in Eq.~\eqref{eq:U} can become small, depending on the evolution time, making $U^{(1)}_{\alpha,\alpha}(t)$ more sensitive to the statistical fluctuations of the correlators, thus reducing the SNR of $S_\alpha(f)$. Note, however, that in general, Eq.~\eqref{eq:auto-PSD} cannot be applied to the cross-PSD $C_{1,2}(f)$ because one attenuation function $\chi_{\alpha,\beta}^{(\pm)}$ is not necessarily much larger than the other when $\tau\gg T_{2}^*$.

Lastly, we state the generalized result that holds without the quasi-static assumption. One obtains it by replacing the prefactor $1/(\pi^2 \tau_\alpha \tau_\beta)$ in Eq.~\eqref{eq:cross_single_shot} with $f^2 / [\sin(\pi f \tau_\alpha) \sin(\pi f \tau_\beta)]$, and similarly for the auto-PSD in Eq.~\eqref{eq:auto-PSD} with $\alpha = \beta$.

\subsection{Choice of probing-sequence parameters}
\label{sec:parameters}

We now discuss how to select the parameters that define the SSCS sequence in Fig.~\ref{fig:sequence}. The experimentally controllable parameters are the average qubit energies $\omega_\alpha$ and the evolution times $\tau_\alpha$ for each qubit $\alpha\in\{1,2\}$. These parameters influence different aspects of the signal: the phases $\phi_\alpha^{if}$ depend on both $\omega_\alpha$ and $\tau_\alpha$, whereas the attenuation functions $\chi_{\alpha,\beta}^{(\pm)}$ are determined solely by $\tau_\alpha$.

\paragraph*{Evolution times.} The optimal choice of evolution times is derived in Appendix~\ref{app:optimal_times}. In brief, the SNR of the extracted spectrum is maximized when $\tau_\alpha\approx T_{2,\alpha}^*$. This holds independently of the strength or character of the noise correlations, and is therefore the recommended choice throughout.

\paragraph*{Qubit energies.} Once the evolution times are fixed, the remaining degree of freedom is the choice of average qubit energies $\omega_\alpha$. These are defined with respect to the rotating frame and set relative to the evolution time. The objective is to ensure that the trigonometric prefactors in Eq.~\eqref{eq:correlators}—specifically $\cos\left(\phi_\alpha^{if}\pm\phi_\beta^{jh}\right)$—do not vanish and are of similar magnitude across correlators, so as to maintain uniform SNR.

For the estimation of cross-PSDs, this condition is satisfied by choosing
\begin{equation}
\omega_1 = \frac{(m + l + 1)\pi}{4\tau_1}, \qquad
\omega_2 = \frac{(m - l)\pi}{4\tau_2},
\label{eq:optimal_cross_frequencies}
\end{equation}
where $m$ and $l$ are integers. These choices guarantee that the two cosine terms accompanying the exponentials of $\chi^{(\pm)}$ in Eq.~\eqref{eq:correlators} are non-zero.

\paragraph*{Auto-PSD considerations.} The estimation of auto-PSDs imposes slightly different constraints. In particular, if Eq.~\eqref{eq:cross_single_shot} is used with $\alpha=\beta$, one must ensure that $\cos(2\phi_\alpha^{if})\neq0$. This is generally not guaranteed under the frequency settings optimal for cross-PSD extraction. Instead, to optimize for auto-PSDs alone, one should choose
\begin{equation}
\omega_\alpha = \frac{m\pi}{2\tau_\alpha},
\label{eq:auto_frequencies}
\end{equation}
with $m$ an integer.

It is possible to select qubit energies that allow for simultaneous estimation of both auto- and cross-PSDs, provided that the non-vanishing cosine conditions above are jointly satisfied. This may require departing from the strict optimal points for either case but can still yield sufficiently high SNR.

\paragraph*{Alternative method for auto-PSD.} As discussed in Sec.~\ref{sec:discusssionOfMainEq} and Eq.~\eqref{eq:auto-PSD}, an alternative method for extracting the auto-PSD involves using long evolution times $\tau_\alpha\gg T_{2,\alpha}^*$. In this limit, the optimal frequency setting is
\begin{equation}
\omega_\alpha = \frac{(2m+1)\pi}{4\tau_\alpha}.
\end{equation}

\section{Validation of the method}

We now proceed to validate the method using both simulated and experimental data. First, we apply it to computer-generated data with known noise characteristics. We then demonstrate its performance on measurements from a spin-qubit device.

\subsection{Computer-simulated data}

We begin by computationally generating time traces of two fluctuating qubit energies $\delta\omega_\alpha(t)$ with set auto-PSDs and cross-PSD, using the methods described in Ref.~\cite{Gutierrez-Rubio2022}. We consider spectra composed of $1/f$ and Lorentzian components, which are features typically found in solid-state qubits. In particular, we choose PSDs going as $C_{\alpha,\beta}(f)=I_{\alpha,\beta}/f+J_{\alpha,\beta}/(1+4\pi^2f^2t_c^2)$, with $I_{1,1}=I_{2,2}=I_{1,2}=10^{11}$~Hz$^2$ and $J_{1,1}=J_{2,2}=-J_{1,2}=10^6$~Hz$^2$/Hz, where the negative sign of $J_{1,2}$ is chosen to induce a phase jump in the cross-PSD, a feature commonly observed in spin-qubit experiments \cite{Yoneda2023,Rojas-Arias2023,Rojas-Arias2025}. We simulate 100 traces of qubit energies with values separated by a time step $\Delta t=250\ \mu$s, and each trace having a length of $N=10^6$, covering a time span of $250$ seconds. {We set the evolution times to be equal to the coherence times $\tau_\alpha=T_{2,\alpha}^*$, since this condition maximizes visibility (see Appendix~\ref{app:optimal_times})}. We choose the qubit energies $ \omega_\alpha $, defined relative to a rotating frame, to satisfy $ \omega_1 = \pi/4\tau_1 $ and $ \omega_2 = 0 $, following Eq.~\eqref{eq:optimal_cross_frequencies} with $m=l=0$.

From the fluctuating frequencies, we generate error-free ($A_\alpha=0$, $B_\alpha=1$) single-shot $\xi_\alpha^{if}$ binary readouts (either $+1$ or $-1$), representing the qubit states, following the quantum-mechanical expectation $P_\alpha^{XX}$ from Eq.~\eqref{eq:expectation} for even measurement times $2n\Delta t$, and $P_\alpha^{XY}$ for odd measurement times $(2n+1)\Delta t$, with $n=0,\ldots,N-1$. Next, we introduce two types of SPAM errors:  a visibility decay, parameterized by the probability $p_e$ for inverting the readout, and a bias, parameterized by the probability $p_b$ to set the readout at $-1$. These parameters translate to the SPAM errors of Eq.~\eqref{eq:average} by $A_\alpha=-p_b$ and $B_\alpha=(1-p_b)(1-2p_e)$.  The procedure above simulates the expected experimental outcomes of the quantum circuit shown in Fig.~\ref{fig:sequence}.

\begin{figure}[tp]
\centering
\includegraphics[width=0.9\columnwidth]{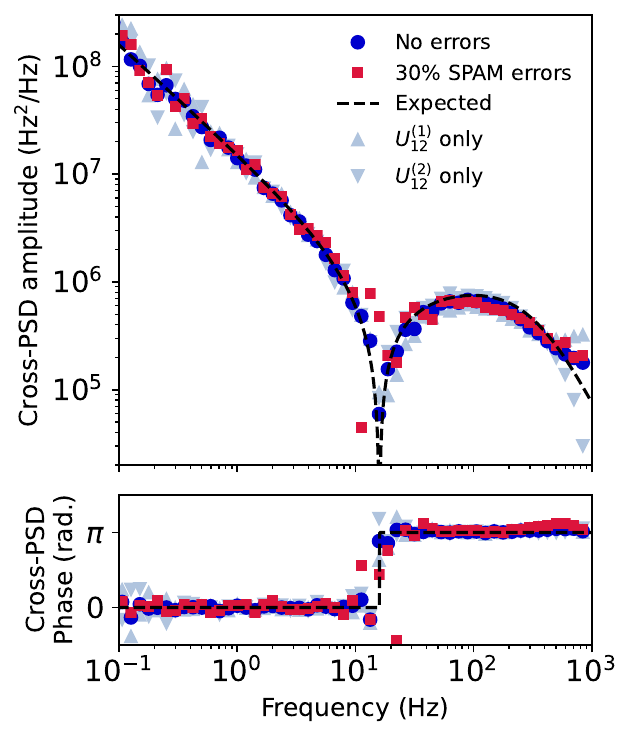}
\caption{
Simulated cross-PSD amplitude and phase of a qubit pair, extracted using our method with (blue circles) and without (red squares) SPAM errors. The blue circles correspond to the average cross-PSD from the two estimators $U^{(a)}$, as described in Appendix~\ref{app:aliasing}. Results obtained by applying Eq.~\eqref{eq:cross_single_shot} to each $U^{(a)}$ individually are shown as light-blue triangles, illustrating the impact of aliasing. The black dashed line represents the reference cross-PSD used to generate the fluctuating qubit energies. For the red squares, SPAM errors are generated with probabilities $p_e=p_b=0.15$.
}
\label{fig:cross-PSD}
\end{figure}

The procedure to obtain the cross-PSD is shown in Fig.~\ref{fig:flow} and goes as follows. From the single-shot readouts we compute the correlators $Q_{1,2}^{if,jh}$, and from them we evaluate $U_{1,2}^{(a)}(t)$ according to their definition in Eq.~\eqref{eq:UV}. As mentioned in Sec.~\ref{sec:discusssionOfMainEq}, due to the alternating fashion in which $R_{XX}$ and $R_{XY}$ are applied in the SSCS sequence given in Fig.~\ref{fig:sequence}, we can only estimate $U^{(1)}_{1,2}$ at even time steps $2n\Delta t$, and $U^{(2)}_{1,2}$ at odd time steps $(2n+1)\Delta t$. We obtain the functions at negative $n$ using the identities $U^{(1)}_{\alpha,\beta}(-t)=U^{(1)}_{\beta,\alpha}(t)$ and $U^{(2)}_{\alpha,\beta}(-t)=U^{(2)}_{\beta,\alpha}(t)+\pi i$. Next, we calculate a discrete Fourier transform of $U_{1,2}^{(a)}$, implementing Eq.~\eqref{eq:cross_single_shot}, resulting in the cross-PSD at frequencies $k/4N\Delta t$ with $k=-N,\ldots,0,\ldots,N-1$. Finally, we average $a=1$ and $2$ to mitigate the aliasing effects (see Appendix~\ref{app:aliasing}). Given that PSDs are typically presented in log-log plots to facilitate the identification of power laws, the linearly spaced points we obtain become exponentially condensed at higher frequencies in the plot and are subject to large statistical fluctuations. To address this, we divide the horizontal axis into equally spaced intervals on a logarithmic scale and plot the arithmetic average of all points within each interval.

The resulting cross-PSD is shown in Fig.~\ref{fig:cross-PSD}. We highlight the following points, which demonstrate that the method is practical. First, the extracted cross-PSD closely matches the expected spectrum, with only a modest reduction in SNR in the presence of SPAM errors, underscoring one of the key strengths of our approach. Second, the effectiveness of aliasing mitigation is evident at higher frequencies: the averaged spectrum from the two estimators $U^{(a)}_{1,2}$ more closely follows the true PSD than either estimator individually, consistent with the analysis in Appendix~\ref{app:aliasing}. Third,  the $\pi$-switch in the correlation phase slightly above 10~Hz is accurately captured, corresponding to the phase jump introduced by the negative cross-term $J_{1,2}$, demonstrating the efficiency of our technique regardless of the specific spectral characteristics of the correlated noise. We provide a Python notebook that implements the full procedure described above, from generating correlated noise to extracting the cross-PSD using SSCS, available at Ref.~\cite{Notebook}.

\begin{figure}[tp]
%\centering
\includegraphics[width=0.9\columnwidth]{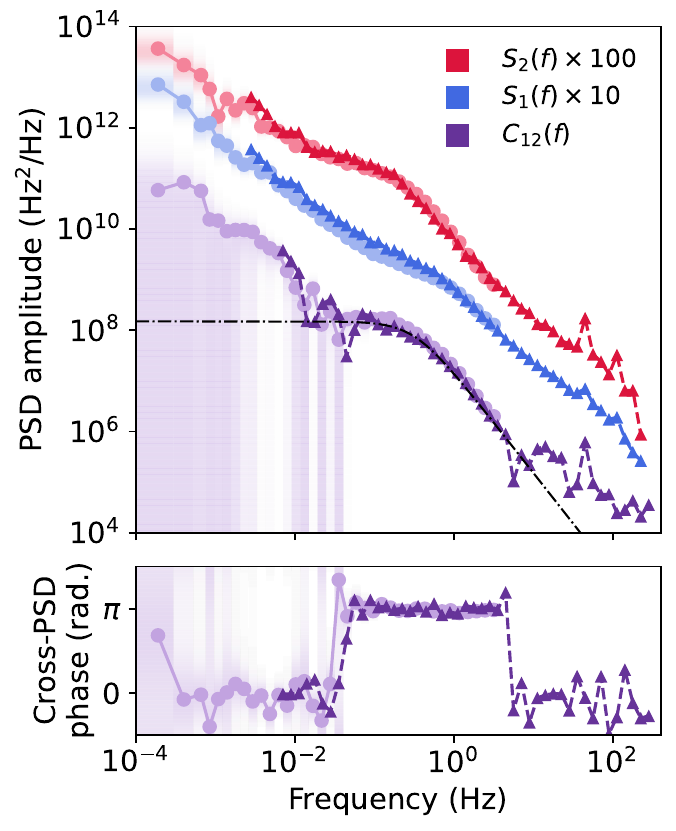}
\caption{Experimental auto-PSDs and cross-PSD of a spin-qubit pair. To improve visibility, $S_1(f)$ is shifted upwards by a factor of 10 and $S_2(f)$ by a factor 100. The PSDs extracted from time traces of the qubit energies (benchmarks) are shown in light-colored circles. The linear color gradients, obtained following the Bayesian estimation of qubit energies of Ref.~\cite{Gutierrez-Rubio2022}, depict a degree of confidence as a probability distribution. The results of SSCS are shown as darker-colored triangles.  The black dashed-dotted line shows a Lorentzian $\propto 1/(1+4\pi^2f^2t_c^2)$ with $t_c=0.5$ s, as a suggestive fit of a part of the cross-PSD.}
\label{fig:experiment}
\end{figure}

\subsection{Experimentally measured data}

We finally apply the protocol to experimental data, as the most stringent test. The data were measured on a pair of spin qubits in the 5-qubit device described in Ref.~\cite{Wu2025}. Details about it are in Appendix~\ref{app:experiment_details}, together with further information on the experiments. Since in this case we neither have the exact spectrum, nor expect it to have any simple shape, we benchmark SSCS with an established method. This is possible for low enough frequencies, where auto- and cross-PSDs can be extracted from estimated qubit energies, as explained in the introduction and (including also cross-PSDs) demonstrated in Refs.~\cite{Yoneda2023, Rojas-Arias2023,Rojas-Arias2025}.

The resulting PSDs are shown in Fig.~\ref{fig:experiment}. The benchmark spectra obtained from time traces of estimated qubit energies are plotted in light-colored circles. While the overall trend is qualitatively a $1/f$ falloff, the spectra have rather rich structure on top of this trend. Furthermore, there is a correlation-phase switch at around 0.04 Hz, moving from positive (phase 0) to negative (phase $\pi$) correlations. This phase shift reflects a change in the dominant noise source in this frequency range, as we explore below.

The PSDs extracted using SSCS are shown in the same figure as darker-colored triangles. These were obtained by following the procedure summarized in Fig.~\ref{fig:flow}. Namely, we first execute the sequences described in Sec.~\ref{sec:sequences}, then we estimate the single-shot correlators as explained in Sec.~\ref{sec:statistics}, and finally obtain the cross-PSD by Eq.~\eqref{eq:cross_single_shot} and auto-PSDs by Eq.~\eqref{eq:auto-PSD}. We observe that first, the method extends the accessible spectral range by two orders of magnitude toward higher frequencies for all quantities. Second, in the frequency range where SSCS overlaps with the benchmark method, all PSD elements match quantitatively, including auto-PSDs, cross-PSD magnitude, and its phase. Altogether, SSCS enables spectral characterization over five decades in frequency on its own, and nearly seven decades when combined with the conventional technique. In our implementation, the accessible range extends up to $\approx500$~Hz, limited by the relatively slow initialization and readout cycle of the five-qubit array. With more optimized hardware cycles, where readout times as short as $2~\mu$s have been demonstrated in Ref.~\cite{Takeda2024}, SSCS should be extendable beyond $10^5$~Hz.

We finish by discussing a few spectral features that are uncovered by the new method. The auto-PSDs of both qubits display noise peaks at 50 Hz and 100 Hz. They correspond to the first and second harmonics of the electrical outlet in Japan. Resolving these peaks (they are less clear for qubit 1, but still discernible) underscores both the method sensitivity and the benefit of extending the spectral range. The cross-PSD also shows the 50 Hz peak with the phase implying positive (in-phase) correlations. In addition to the phase switch at 0.04 Hz mentioned above, the cross-PSD phase displays a second $\pi$-switch around 5 Hz. As the correlation phase is related to the location of noise sources~\cite{Rojas-Arias2025}, detecting such phase changes provides hints on the spatial configuration of the noise environment. Specifically here, the good fit of the cross-PSD to a Lorentzian (black curve) in the range where correlations are negative suggests the presence of a two-level fluctuator located between the qubits, with a characteristic switching time of about half a second. 

\section{Conclusions}

We have developed a practical method for extracting auto- and cross-power spectral densities of dephasing noise in qubit pairs. It requires only basic single-qubit Ramsey-type sequences with single-shot readouts, making it straightforward to implement on most qubit platforms. Our method enables noise spectroscopy over a wide frequency range set only by the experiment repetition rate. It is resilient to SPAM errors (they need neither to be known nor calibrated) and aliasing effects, and we demonstrate its reliability on both simulated and experimental data obtained from a five-qubit silicon spin device. To support broader adoption, we provide a Python notebook implementation in Ref.~\cite{Notebook}.
By extending the accessible frequency range of noise spectroscopy and enabling the direct extraction of noise correlations, this method contributes to a more complete understanding of dephasing noise in multi-qubit systems.

\acknowledgments
We would like to acknowledge Kenta Takeda, Takashi Nakajima, and Takashi Kobayashi for their assistance with the technologies used in our experiments, Akito Noiri for the fabrication of the device, and Giordano Scappucci for providing the Si/SiGe heterostructure.
We thank for the financial support from the Swiss National Science Foundation, NCCR SPIN grant No. 51NF40-180604,  MEXT Quantum Leap Flagship Program (MEXT Q-LEAP) grant number JPMXS0118069228, and JST Moonshot R\&D grant number JPMJMS226B. J.S.R.-A. and L.C.C. acknowledge support from the Gutaiteki Collaboration Seed. Y.-H.W. acknowledges support from RIKEN's IPA program.
%\clearpage

\appendix

\section{Effects of aliasing}\label{app:aliasing}

Noise spectra in solid-state systems often have a $1/f$-like shape. This means that, independent of the signal sampling rate, there will always remain an appreciable high-frequency tail of noise beyond the measurement Nyquist frequency~\cite{Kirchner2005}. The Fourier transform, as a typical method to characterize the noise spectral components, is then affected by aliasing. The aliasing is an undesired effect where the noise power of frequencies higher than the Nyquist frequency $f_N$ is folded over to the fundamental frequency range $[0,f_N]$.

Consider a continuous time-dependent variable $x(t)$ that has a Fourier Transform $X(f)=\int_{-\infty}^\infty dt\ x(t) \exp(2\pi i f t)$. For us, $x(t)$ is a cross-correlator of qubit energies and $X(f)$ is its associated cross-PSD. Let us consider $x(t)$ sampled at discrete times $2 n \Delta t$, with $n$ an integer. It corresponds to the case of $U^{(1)}_{\alpha,\beta}$ in the main text. The Fourier transform $Y_e(f)$ of the infinite number of delta peaks $x(2n\Delta t)$ is~\cite{Kirchner2005}
\begin{align}
Y_e(f)=X(f)+\sum_{k=1}^\infty\left[X(k f_s+f)+X^*(k f_s-f)\right],
\label{eq:aliasing}
\end{align}
with $f_s=1/2 \Delta t$, the sampling rate. Equation~\eqref{eq:aliasing} displays the aliasing: The obtained Fourier transform at frequency $f$ contains, apart from the first term on the right-hand side being the desired Fourier transform $X$, also an infinite sum of contributions at larger frequencies, the ones shifted by integer multiples of $f_s$. 

Next, let us consider that the quantity $x(t)$ is sampled at odd time steps $(2n+1)\Delta t$, corresponding to $U^{(2)}_{\alpha,\beta}$ in the main text. Its Fourier transform $Y_o(f)$ follows as 
\begin{align}
Y_o(f)=X(f)+\sum_{k=1}^\infty(-1)^k\left[X(k f_s+f)+X^*(k f_s-f)\right].
\label{eq:aliasing2}
\end{align}
The result is analogous to Eq.~\eqref{eq:aliasing}, except for a minus sign at frequencies folded over by odd multiples of $f_s$. Due to the sign flip, those aliased frequencies drop out in the average $\bar{Y}\equiv(Y_e+Y_o)/2$, 
\begin{align}
\bar{Y}(f)=X(f)+\sum_{k=1}^\infty\left[X(2k f_s+f)+X^*(2k f_s-f)\right].
\label{eq:aliasing_average}
\end{align}
As a result, the averaged Fourier transform $\bar{Y}(f)$ is less affected by aliasing since half of the aliased contributions are removed. In fact, Eq.~\eqref{eq:aliasing_average} corresponds to Eq.~\eqref{eq:aliasing} with a double sampling rate. This is to be expected, as we could combine both $x(2n\Delta t)$ and $x((2n+1)\Delta t)$ into a single trace $x(n\Delta t)$, with half the time step, before applying the Fourier transform. This would yield a factor of 2 increase in the fundamental frequency range. However, in the main text $U^{(1)}_{\alpha,\beta}(2n\Delta t)$ and $U^{(2)}_{\alpha,\beta}((2n+1)\Delta t)$ differ by a constant, which forbids us from combining them into a single time trace without creating artifacts in the Fourier transform. 

\begin{figure}[htbp]
\centering
\includegraphics[width=0.9\columnwidth]{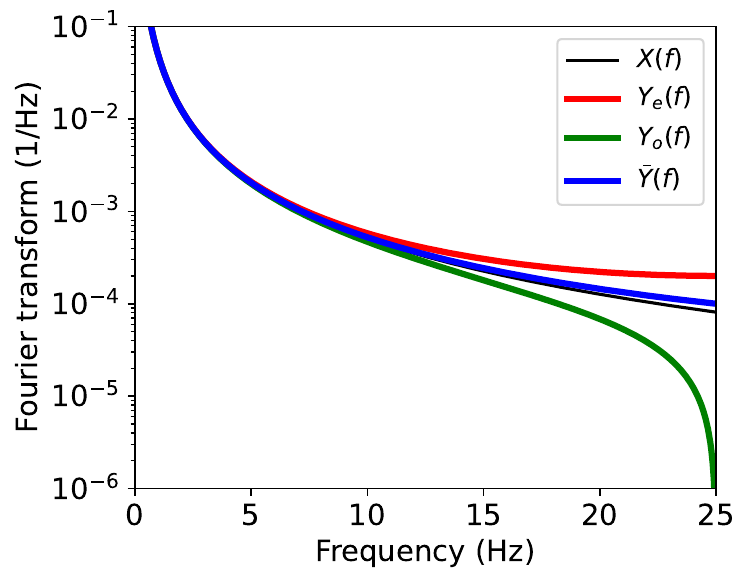}
\caption{Comparison of the effect of aliasing in the Fourier transform of $x(t)$, when sampled at odd (green) and even (red) time steps. The significantly less aliased curve (blue) corresponds to averaging the results of both time steps.
}
\label{fig:alias}
\end{figure}

We finish with an example that highlights the effect of aliasing. Consider the function $x(t)=\exp(-|t|/t_0)$ sampled at even and odd multiples of $\Delta t$. While the Fourier transform of this function is not $1/f$, it corresponds to noise induced by a two-level fluctuator as a plausible noise source for solid-state qubits, and has the advantage that it can be solved analytically. We get
\begin{subequations}
\begin{align}
X(f)&=\frac{2t_0}{1+4\pi f^2t_0^2},\\
Y_e(f)&=X(f)\frac{(2\pi f \Delta t)^2}{\sin^2\left(2\pi f \Delta t\right)},\\
Y_o(f)&=X(f)\frac{(2\pi f \Delta t)^2\cos\left(2\pi f \Delta t\right)}{\sin^2\left(2\pi f \Delta t\right)},\\
\bar{Y}(f)&=X(f)\frac{(\pi f \Delta t)^2}{\sin^2\left(\pi f \Delta t\right)}.
\end{align}
\label{eq:ft}
\end{subequations}
The formulas from Eq.~\eqref{eq:ft} are plotted in Fig.~\ref{fig:alias}, for $t_0=1$ s and $\Delta t=0.01$ s. One can see how aliasing amplifies as the frequency approaches the Nyquist frequency $f_N=f_s/2$. Specifically, $Y_e(f_N)=\frac{\pi^2}{4}X(f_N)$ and $Y_o(f_N)=0$, while $\bar{Y}(f_N)=\frac{\pi^2}{8}X(f_N)$. It is interesting to note the drastic difference that aliasing has for sampling at even or odd time steps. Sampling at even multiples of $\Delta t$ makes the Fourier transform take more than twice the true value at the Nyquist frequency, while sampling at odd multiples of $\Delta t$ causes a sharp decay towards 0. That is a general behavior valid beyond the particular example used here. One can prove that for $x(t)$ real and even function of $t$, $Y_e(f_N)\geq 2X(f_N)$ and $Y_o(f_N)=0$, while $\bar{Y}(f_N)\geq X(f_N)$. It is clear that $\bar{Y}$ is closer to the true Fourier transform $X$ than either $Y_o$ or $Y_e$, and we thus adopt it as our estimate of $X$.

\section{Optimal evolution time to enhance visibility of cross-PSDs}\label{app:optimal_times}

The signal-to-noise ratio (SNR) of the spectra obtained from the SSCS protocol depends critically on how well we can reconstruct the functions $U^{(a)}_{\alpha,\beta}$ from measured single-shot data. A key factor in this reconstruction is the choice of evolution times $\tau_{\alpha,\beta}$, which can strongly affect the visibility of the correlators. In this section, we derive the optimal evolution time. We focus on the function $U^{(1)}_{\alpha,\beta}$, but the procedure is analogous for $U^{(2)}_{\alpha,\beta}$ and yields the same values for $\tau_{\alpha,\beta}$.

The function $U^{(1)}_{\alpha,\beta}(t)$ is obtained from linear combinations of the single-shot correlators $G_{\alpha,\beta}^{(\pm)}(t) \equiv Q_{\alpha,\beta}^{XX}(t) \pm Q_{\alpha,\beta}^{YY}(t)$. To achieve good SNR, the values of $G_{\alpha,\beta}^{(\pm)}(t)$ should be well above the statistical noise. From Eqs.~\eqref{eq:correlators} and \eqref{eq:envelope}, one can see that if the evolution times are too long, the functions are exponentially small, $\lim_{{\tau_{\alpha,\beta} \rightarrow \infty}}G_{\alpha,\beta}^{(\pm)}(t)\rightarrow 0$, resulting in vanishing contrast and no information. Conversely, if the evolution times are too short, $\lim_{{\tau_{\alpha,\beta} \rightarrow 0}}G_{\alpha,\beta}^{(\pm)}(t) \rightarrow 0$ as well, since the qubits have not evolved long enough to accumulate appreciable noise to resolve $\chi^{(+)}$ from $\chi^{(-)}$, see Eq.~\eqref{eq:envelope}. Thus, we seek an intermediate regime where the qubits are sufficiently affected by noise to encode useful information, while retaining enough coherence to read it out.
To identify this regime, we express $G_{\alpha,\beta}^{(\pm)}$ as:
\begin{align}
&G_{\alpha,\beta}^{(\pm)}(\tau_\alpha,\tau_\beta;t) = B_\alpha B_\beta \cos\left(\omega_\alpha\tau_\alpha \mp \omega_\beta\tau_\beta\right) \times \nonumber\\
&\qquad \times \left[e^{-\frac{1}{2}\left(x_\alpha + x_\beta \mp 2\sqrt{x_\alpha x_\beta} \rho_{\alpha,\beta}(t)\right)} - e^{-\frac{1}{2}(x_\alpha + x_\beta)}\right],
\end{align}
where we define $x_\alpha \equiv \braket{\delta\omega_\alpha^2} \tau_\alpha^2$ and the normalized time-domain correlation coefficient
\begin{align}
\rho_{\alpha,\beta}(t) \equiv \frac{\braket{\delta\omega_\alpha(t') \delta\omega_\beta(t'+t)}}{\sqrt{\braket{\delta\omega_\alpha^2}\braket{\delta\omega_\beta^2}}}.
\end{align}
Since $\rho_{\alpha,\beta}(t)$ can range between –1 and 1 depending on the details of the correlated noise---which are not known a priori---we avoid optimizing $G_{\alpha,\beta}^{(\pm)}$ for a specific value of $\rho_{\alpha,\beta}$. Instead, we average over all possible values of $\rho_{\alpha,\beta}$ to obtain a general measure of signal strength:
\begin{align}
g_{\alpha,\beta}^{(\pm)} &\equiv \frac{1}{2} \int_{-1}^{1} \mathrm{d}\rho_{\alpha,\beta}\ G_{\alpha,\beta}^{(\pm)}\nonumber\\
& = B_\alpha B_\beta \cos\left(\omega_\alpha\tau_\alpha \mp \omega_\beta\tau_\beta\right) \times \label{eq:average_w} \\
&\qquad \times \left[\frac{e^{-\frac{1}{2}(x_\alpha + x_\beta)}\left(\sinh\left(\sqrt{x_\alpha x_\beta}\right) - \sqrt{x_\alpha x_\beta}\right)}{\sqrt{x_\alpha x_\beta}}\right].\nonumber
\end{align}
To find the optimal evolution times, we maximize $g_{\alpha,\beta}^{(\pm)}$ with respect to $x_\alpha$ and $x_\beta$ under the constraint $x_{\alpha,\beta} \geq 0$. The cosine term in Eq.~\eqref{eq:average_w} does not influence the maximization since $\omega_{\alpha,\beta}$ are typically set such that $\omega_\alpha \tau_\alpha$ and $\omega_\beta \tau_\beta$ are constant phases. This leaves the bracketed term as the relevant quantity for optimization. A numerical maximization yields the optimal value $x_\alpha = x_\beta = 2.72877$. It corresponds to the evolution times
\begin{align}
\tau_{\alpha} = \sqrt{\frac{2.72877}{\braket{\delta\omega_\alpha^2}}} = 1.16807\times T_{2,\alpha}^*,
\end{align}
and similarly for $\tau_\beta$. For simplicity, we suggest to use $\tau_\alpha \approx T_{2,\alpha}^*$ and $\tau_\beta \approx T_{2,\beta}^*$, expecting no relevant difference, especially since in the non-ergodic regime $T_2^*$ fluctuates.

\section{Experimental details}\label{app:experiment_details}

We work with the spin-qubit device reported in Ref.~\cite{Wu2025}. The device consists of a linear five-quantum dot array fabricated on top of a $^{28}$Si/SiGe heterostructure with a residual concentration of 800ppm $^{29}$Si isotopes in the quantum well. Specifics on the setup, operation, and readout protocols can be found in Ref.~\cite{Wu2025}. While qubits are defined across the whole array, for this experiment we only work with qubits 2 and 3.

First, we perform noise spectroscopy of the qubit pair by performing an interleaved Ramsey experiment as described in Ref.~\cite{Rojas-Arias2023}. During the experiment, we collect records in which the free-evolution time is changed from 0 to $4\ \mu$s in $0.04\ \mu$s steps. One value of each qubit energy is estimated from a single record by Bayesian estimation~\cite{Delbecq2016,Nakajima2020}. An interleaved Ramsey cycle takes $t_{\mathrm{cycle}}=1.580\ $ms, so the estimated qubit energies have a time step $\Delta t = 100\times t_{\mathrm{cycle}}= 0.158\ $s. We acquire a total of $2\times10^5$ records such that the time traces of the qubit energies cover a total time of 8.77 hours. From these time traces, we calculate the auto- and cross-PSDs shown as circles in Fig.~\ref{fig:experiment} using the Bayesian estimation of correlation functions from Ref.~\cite{Gutierrez-Rubio2022}. The corresponding probability distributions of the estimations are shown as linear color gradients.

Next, we implement the SSCS protocol described in Fig.~\ref{fig:sequence} with two different modes. In the first mode, with the goal of extracting the cross-PSD of the qubit pair following Eq.~\eqref{eq:cross_single_shot}, we set the evolution times to $\tau_1=\tau_2=5\ \mu$s,  approximately matching $T_2^*$ for both qubits. The detunings of the qubit energies relative to their respective rotating frame frequencies are $\omega_1 = 0$ and $\omega_2 = \pi / (3\tau_2)$. The cycle time for each subsequence $R_{XX}$ and $R_{XY}$ is $\Delta t=791\ \mu$s, limited by the initialization and readout of the whole array. We repeat the $R_{XX}$–$R_{XY}$ subsequence $10^6$ times, covering a total time of 26.37 minutes, which we define as a single batch. From each batch, we obtain a cross-PSD $C_{1,2}(f)$. To improve the SNR, we acquire a total of 8 batches, and their average is presented as purple triangles in Fig.~\ref{fig:experiment}.

In the second mode, meant to extract the auto-PSDs of the qubits following Eq.~\eqref{eq:auto-PSD}, we set the evolution times $\tau_1=\tau_2=7\ \mu$s such that $\tau_1,\tau_2>T_2^*$. The average qubit energy detunings in the rotating frame are $\omega_1=\pi/4\tau_1$ and $\omega_2=\pi/3\tau_2$. Each subsequence takes $\Delta t = 1.062$~ms and is applied $10^6$ times, covering a total time of 35.40 minutes. We acquire 10 batches and the average auto-PSDs $S_1(f)$ and $S_2(f)$ from them are shown as blue and red triangles in Fig.~\ref{fig:experiment}, respectively.

\bibliography{references.bib}
\end{document}